\documentclass{PoS}

\leftline{\parbox{3cm}{\large\rm DESY 06-192 \\ HU-EP-06/37 \\ MS-TP-06-12}
\vspace{1cm}}

\title{Twisted mass QCD thermodynamics: \\first results on apeNEXT}

\ShortTitle{Twisted mass QCD thermodynamics: first results on apeNEXT}

\author{
Ernst-Michael Ilgenfritz, Michael M\"uller-Preussker and  Andre Sternbeck
\footnote{Address since Sept.1 2006: CSSM, School of Chemistry \& Physics
   The University of Adelaide, AUSTRALIA 5005}
\\
Humboldt-Universit\"at zu Berlin, Institut f\"ur Physik,
Newtonstr. 15 , 12489 Berlin\\
        E-mail: \email{ilgenfri@physik.hu-berlin.de},
        \email{mmp@physik.hu-berlin.de},
        \email{andre.sternbeck@adelaide.edu.au} 

}
\author{Karl Jansen and Ines Wetzorke \\
NIC, Platanenallee 6, 15378 Zeuthen \\
E-mail: \email{Karl.Jansen@desy.de},\email{Ines.Wetzorke@desy.de}}
\author{\speaker{Maria Paola Lombardo} \\
Istituto Nazionale di Fisica Nucleare, LNF, Via Enrico Fermi 40, I 00044, 
Frascati (Roma)  \\
        E-mail: \email{lombardo@lnf.infn.it}}
\author{Owe Philipsen \\
Westf\"alische Wilhelms-Universit\"at M\"unster, 
Institut f\"ur Theoretische Physik, 
Wilhelm-Klemm-Str.9, 48149 M\"unster \\
E-mail: \email{ophil@uni-muenster.de}}

\abstract{The motivations for simulating QCD thermodynamics with 
Wilson fermions and a twisted mass term are introduced. 
The twisted mass approach provides a natural infrared cutoff and 
$O(a)$ improvement at maximal twist, and can be extended to finite temperature. 
Our strategy for exploring the QCD phase diagram at finite temperature in 
this setup, while taking advantage of the results at $T=0$, is explained. 
The first results for the order parameters and susceptibilities 
on a $16^3 \times 8$ lattice are presented. 
All dynamical simulations are carried out on the apeNEXT facility in Rome.
 }

\FullConference{XXIVth International Symposium on Lattice Field Theory\\
                July 23-28, 2006\\
                Tucson, Arizona, USA}

\begin{document}

\section{Motivations} 

Ultimately, we would like to study the finite temperature phase transition 
in QCD and the properties of the plasma phase (QGP), with physical values 
of the quark masses and in the continuum limit. 
The present project attempts to set the stage for such a complete study.

In this study we consider QCD with two degenerate flavors.
A cross check of the lattice results
obtained with staggered and Wilson fermion action will help 
controlling discretization artifacts, while a twisted mass term 
is expected to facilitate the continuum and chiral limits.

Consider the phase diagram of two-flavor QCD in the temperature-mass 
plane:  
the first order deconfinement transition stemming from the infinite mass
(or quenched) theory weakens with lower quark masses,
until it turns into a smooth crossover for intermediate quark masses.
In the chiral limit there has to be a true phase transition again,
but its nature is still under investigation~\cite{Philipsen:2005mj}.

As a first step of our program, we wish to find the location of the
phase boundary between hadronic and plasma phase, {\it i.e.} the 
pseudo-critical temperature and mass combinations, while taking advantage 
of the properties of twisted mass QCD. This means that 
we will have to explore a three-dimensional space of temperature T, 
bare quark mass m, and twisted mass parameter $\mu$.

\section{Why Twisted Mass QCD Thermodynamics ?}

Wilson fermions have several advantages over staggered fermions, 
but they also have a more subtle chiral behavior, and a complicated 
phase structure, both at $T=0$~\cite{Ilgenfritz:2003gw,Farchioni:2004us}
and at finite 
temperature~\cite{AliKhan:2000iz,AliKhan:2001ek,Creutz:1996bg,Ilgenfritz:2005ba}.
The twisted mass approach
improves over the standard Wilson behavior in two ways: 
first, it prevents the occurrence of exceptional configurations and 
should make it relatively easy to reach mass values of the light 
pseudoscalar mesons close to the physical pion mass; second, once 
the Wilson hopping parameter $\kappa$ is set to its critical value, 
the twisted mass term behaves as a conventional quark mass, and, at 
the same time, an $O(a)$ improvement is automatically guaranteed.
For recent results see Refs.~\cite{Farchioni:2005ec,Jansen:2006rf} and for
a review Ref.~\cite{Shindler:2005vj}.

In this first report we search for the transitions between the hadronic 
and plasma regimes by varying the Wilson hopping parameter $\kappa$ 
related to the bare quark mass by $\kappa=1/(2m+8)$ 
at fixed $\beta$ and fixed twisted mass parameter $\mu$. 

\section{Strategy and simulations} 

The simulations were performed on a $16^3 \times 8$ lattice with 
an improved version of the HMC algorithm as detailed in Ref.~\cite{Urbach:2005ji}
and with a Symanzik tree-level improved gauge action. They used approximatively
three months$\times$crate of apeNEXT~\cite{Belletti:2006nw}. 
We choose to work at $\beta = 3.75$ and $\beta = 3.9$ 
in order to take advantage of the $T = 0$ 
results~\cite{Farchioni:2005ec,Jansen:2006rf}. In principle, we can then cross 
the (pseudo-)critical line at a fixed temperature by tuning the quark mass, 
either by varying $\kappa$ and/or $\mu$.

For this strategy to be successful, the simulation parameters $N_t$ and $\beta$ 
need to satisfy:
\begin{equation}
T_c^{chiral} < T^{simulation} = 1/(N_t a(\beta)) < T_c^{quenched} \; .
\end{equation}
For $T^{simulation} > T_c^{quenched}$ the hadronic phase cannot exist,
while for $T^{simulation} < T_c^{chiral}$ the QGP cannot exist, 
irrespective of the mass value.

We fixed $N_t = 8$ by taking into account the lattice spacing from
the $T=0$ studies, namely $a(3.75) \simeq 0.12$ fm, $a(3.9) \simeq 0.095$ fm,
as well as the known critical temperatures,
$T^c_{chiral} = 170$ MeV and $T^c_{quenched} = 270$ MeV.

In the first set of runs reported here we fixed $\mu = 0.005$ and
varied the hopping parameter $\kappa$. 

\section{Results at $\beta = 3.75$}  

\begin{figure}[t]
\includegraphics[width=7.0 truecm,height=5.5 truecm]{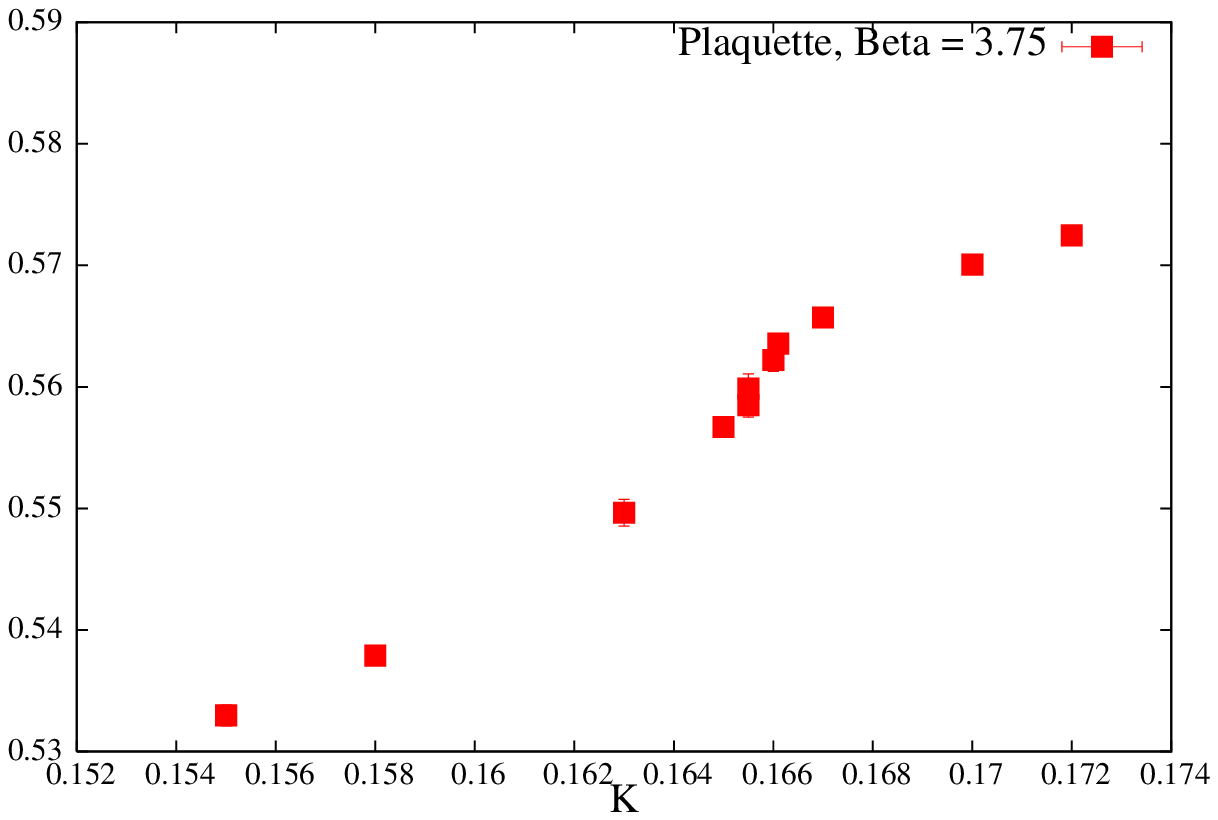}
\hspace*{0.7cm}
\includegraphics[width=7.0 truecm,height=5.5 truecm]{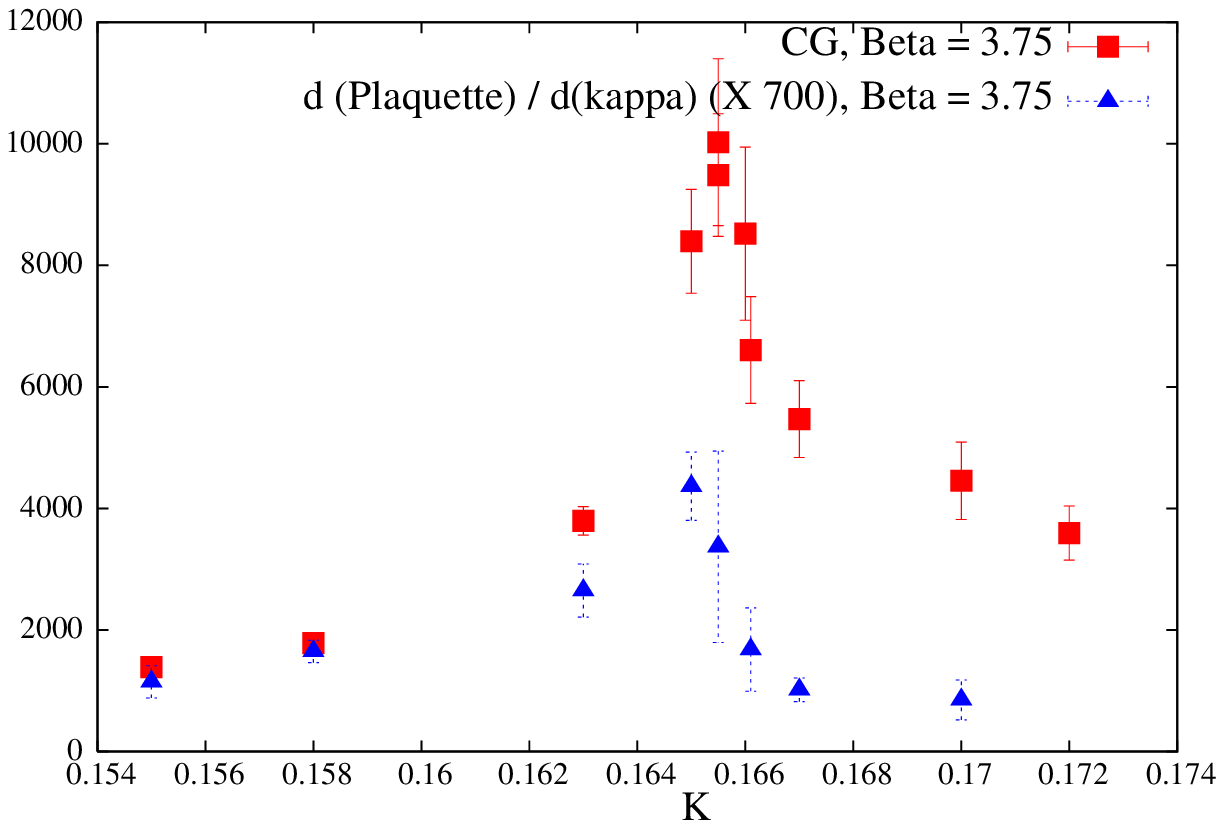}
\caption{\label{plaq} $\langle P \rangle$ as a function of $\kappa$ (left diagram); 
Conjugate Gradient (CG) Iterations superimposed with 
$\frac{\partial \langle P \rangle}{\partial \kappa}$ (magnified) 
as a function of $\kappa$ at fixed  $\beta = 3.75$ and $\mu = 0.005$ 
(right diagram). The results indicate a thermal transition or crossover 
at $\kappa_t = 0.165(1)$.}
\end{figure}

At $\beta = 3.75$, the $T = 0$ results show that the minimum pion mass
which can be reached with our twisted mass parameter $\mu = 0.005$ 
is $m_\pi \simeq 400 {\rm MeV}$ 
extrapolating existing results at $\mu = 0.005$~\cite{Farchioni:2005ec}.
Our first goal here is merely to check whether a thermal phase transition 
or a crossover can be found in the required range $\kappa_t < \kappa_c$.

Figure \ref{plaq} (left) shows a scan of the average plaquette as a function 
of $\kappa$.
The steepest slope of the plaquette as well as the increase of the number 
of CG iterations needed for the inversion shown in Figure \ref{plaq} (right), 
both suggest a crossover or phase transition at~\cite{Farchioni:2005ec}
\begin{equation}
\kappa_t(\beta = 3.75, \mu = 0.005) = 0.165(1) \; . 
\end{equation}
Hence $\kappa_t< \kappa_c(T = 0) = 0.1669$, as required.

\section{Results at $\beta$ = 3.9}  

After the exploratory study at $\beta = 3.75$, the choice $\beta = 3.9$ brings 
us closer to the continuum limit.
Results for $T = 0$ at this $\beta$ are available at a number of values for
the twisted mass parameter $\mu$, see Ref.~\cite{Jansen:2006rf}.
The minimum pion mass at $T = 0$
for our $\mu = 0.005$, inferred from these results, is about 350 MeV.

As a direct comparison between the results at the two $\beta$ values 
we show in Figure 2 the  number of Conjugate Gradient iterations 
required for the  inversion -- which is an  indicator of criticality  --
as a function of $\kappa$.

\begin{figure}[t]
\begin{center}
\includegraphics[width=8.0 truecm,height=4.5 truecm]{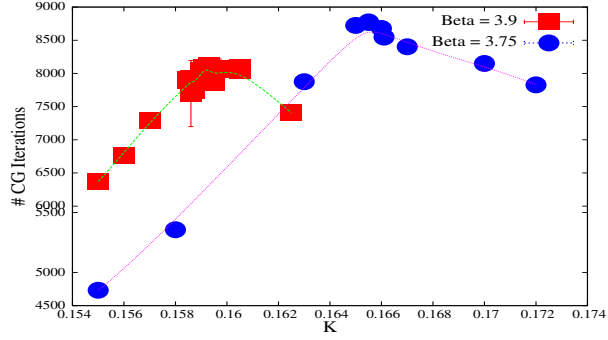}
\caption{The number of Conjugate Gradient iterations as a function of $\kappa$
for the two $\beta$-values.
The solid lines are a smooth interpolation to guide the eye.}
\vspace*{-0.5cm}
\end{center}
\end{figure}

Figure 3 (left) shows a collection of results for the 
expectation value of the plaquette $\langle P \rangle$. 
The errors are smaller than the symbol, the solid line is a Bezier interpolation 
to guide the eye. We performed local fits to a straight line 
$\langle P \rangle = a + b \kappa $ within subsequent 
intervals of width $\Delta \kappa = 0.002$, and we show in the 
same diagram the tangent in the inflection point. 
The parameters $b$ are used as estimators of 
the derivative of the Plaquette w.r.t. to $\kappa$ and are shown in
Figure 3 (right). These results indicate a phase transition or crossover
around $\kappa = 0.1597$ 
located according to the maximal slope $b_{max}$.
\begin{figure}[htb]
\includegraphics[width=10.1 truecm,height=5.5 truecm]{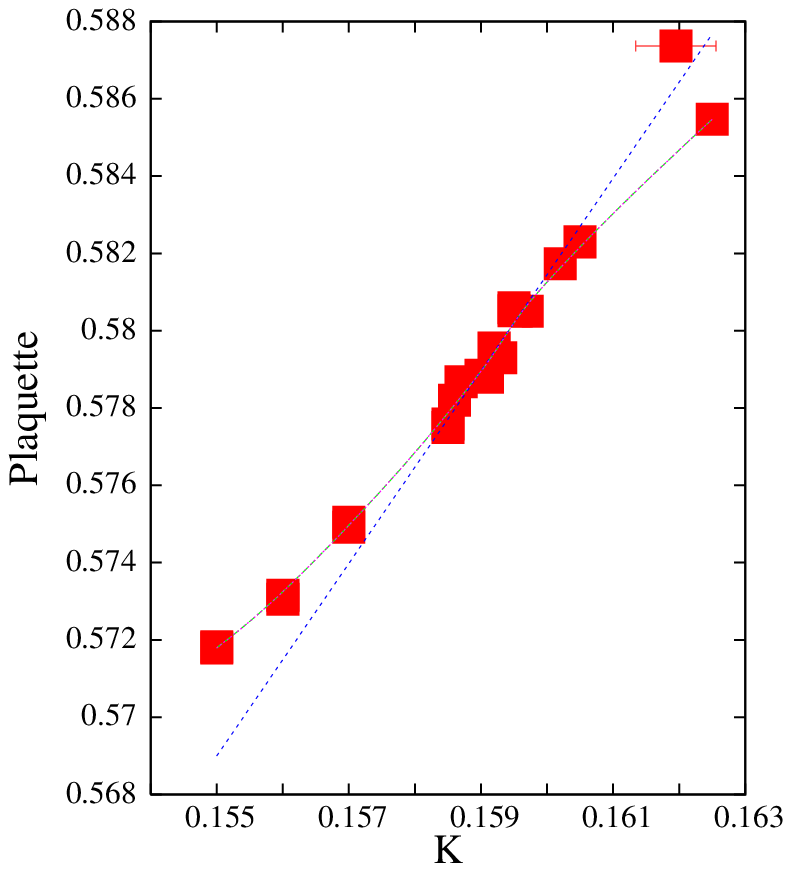}
\hskip -2.0 truecm 
\includegraphics[width=10.1 truecm,height=5.5 truecm]{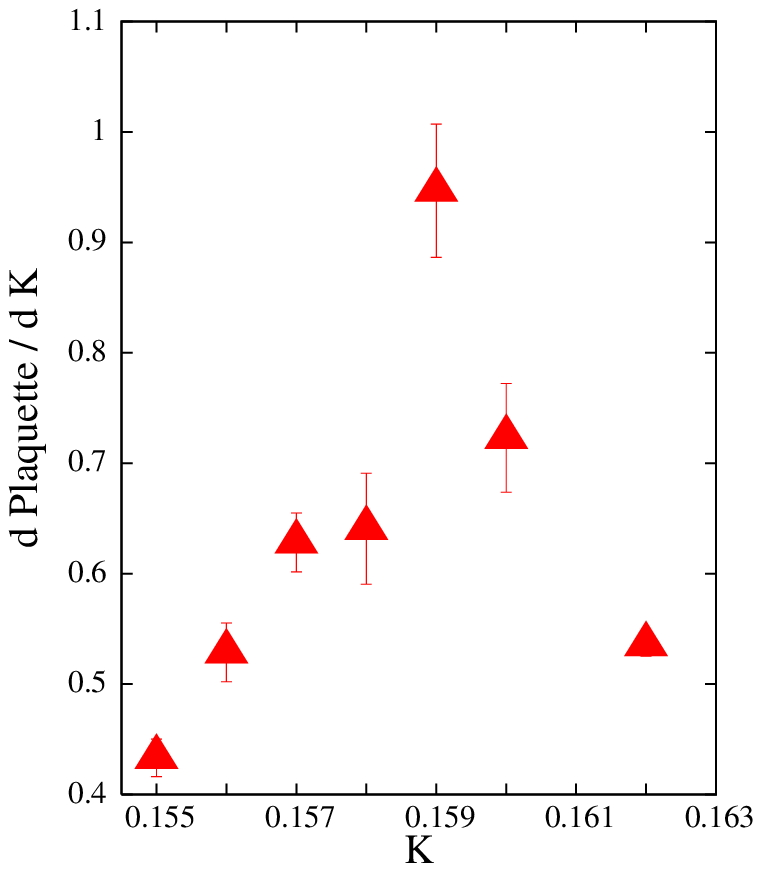}
\caption{$\langle P \rangle$ as a function of $\kappa$ (left diagram); 
$\frac{\partial \langle P \rangle}{\partial \kappa}$  
as function of $\kappa$ at fixed  $\beta = 3.9$ and $\mu = 0.005$ (right diagram)}
\end{figure}
To make this prediction more quantitative, 
our statistics was enhanced to $O(10000)$ HMC trajectories 
on a selected sample of points: 
$\kappa = 0.1586, 0.1591, 0.1593, 0.1597$ in the candidate critical region at 
$\beta = 3.9$. The results (Figure 4) suggest a long autocorrelation 
time  in the critical region.
Given these autocorrelations, our results are very preliminary. Most probably
the errors are underestimated, but still the plots may serve as indicators for
the location of a crossover or transition.
\begin{figure}[t]
\includegraphics[width=7.0 truecm,height=4.5 truecm]{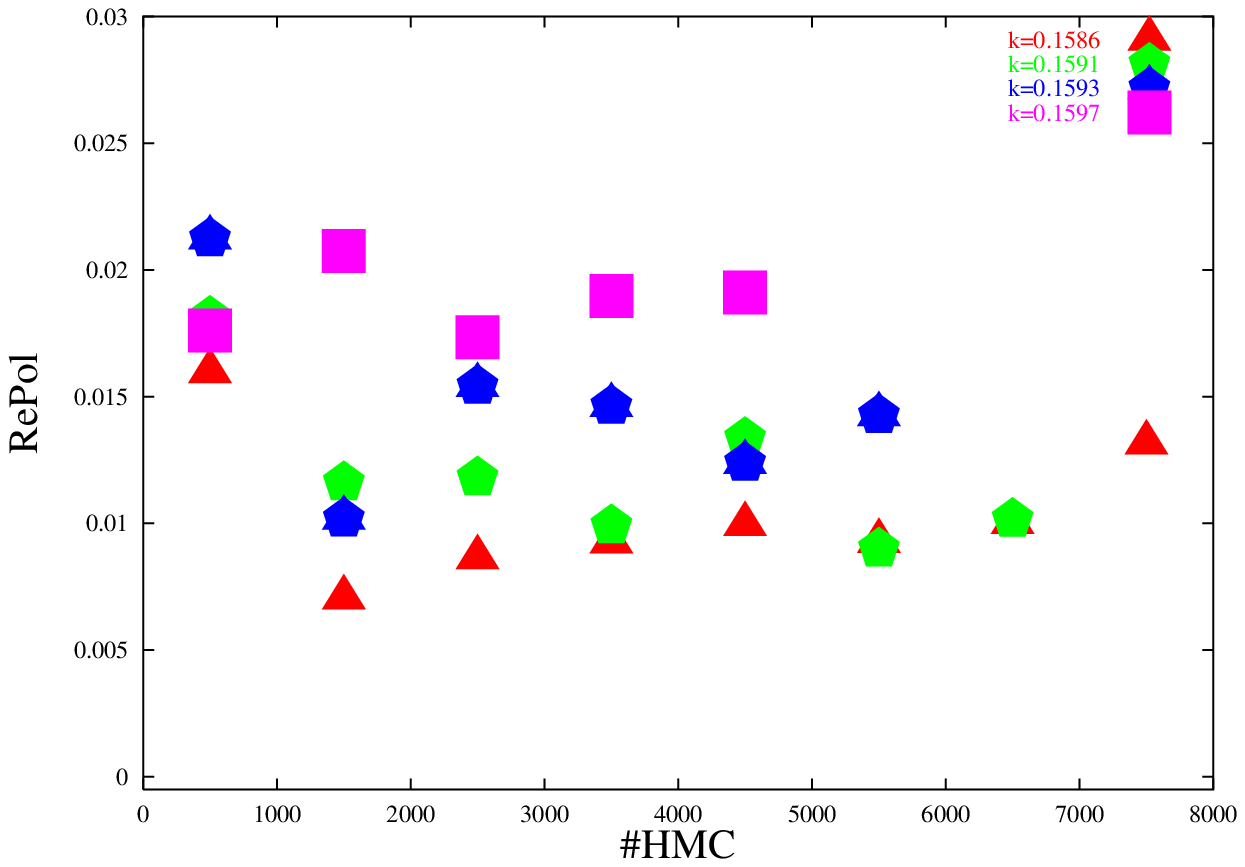}
\hspace*{0.5cm}
\includegraphics[width=7.0 truecm,height=4.5 truecm]{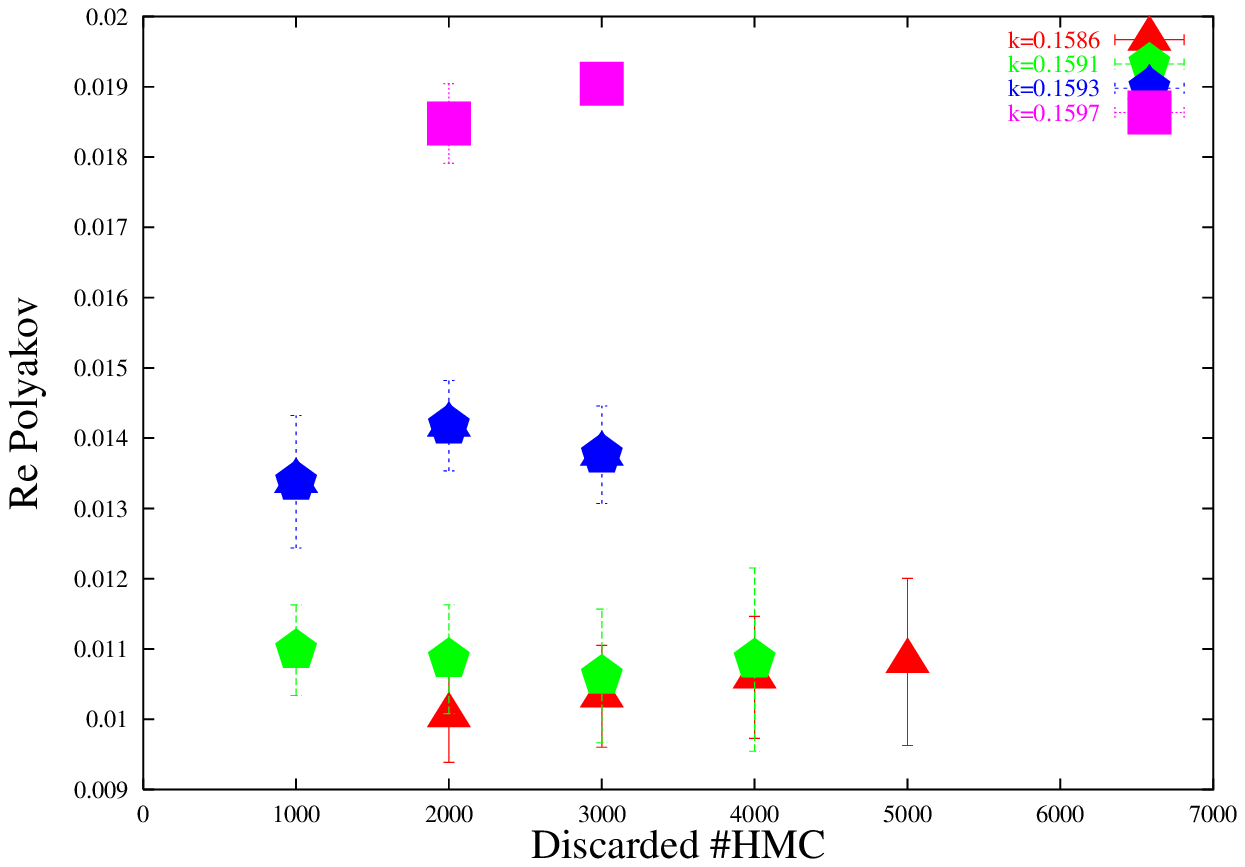}
\caption{HMC evolution and error analysis for the high statistics runs
at $\beta = 3.9$ : binned averages (left) as a function of the 
HMC trajectories, and results and errors as a function
of the discarded HMC trajectories (right).}
\end{figure}

Figure 5 shows the results for the Polyakov loop:
the steepening is clearly seen, mostly thanks to the latest, high statistics
results.
The Polyakov loop increases in the same $\kappa$ interval as the one observed for 
the plaquette.
The Polyakov loop histograms of the HMC results after thermalization 
are narrow in the two 
pure phases, and broaden around at $\kappa=0.1597$, 
indicating an increase of the fluctuations and a critical behavior. 
We see no two-state signal (two peaks in the histograms), 
which seems to exclude a first order transition, 
but only a finite size analysis can assess 
with confidence the nature of the critical behavior.
\begin{figure}[b]
\includegraphics[width=7.0 truecm,height=5.5 truecm]{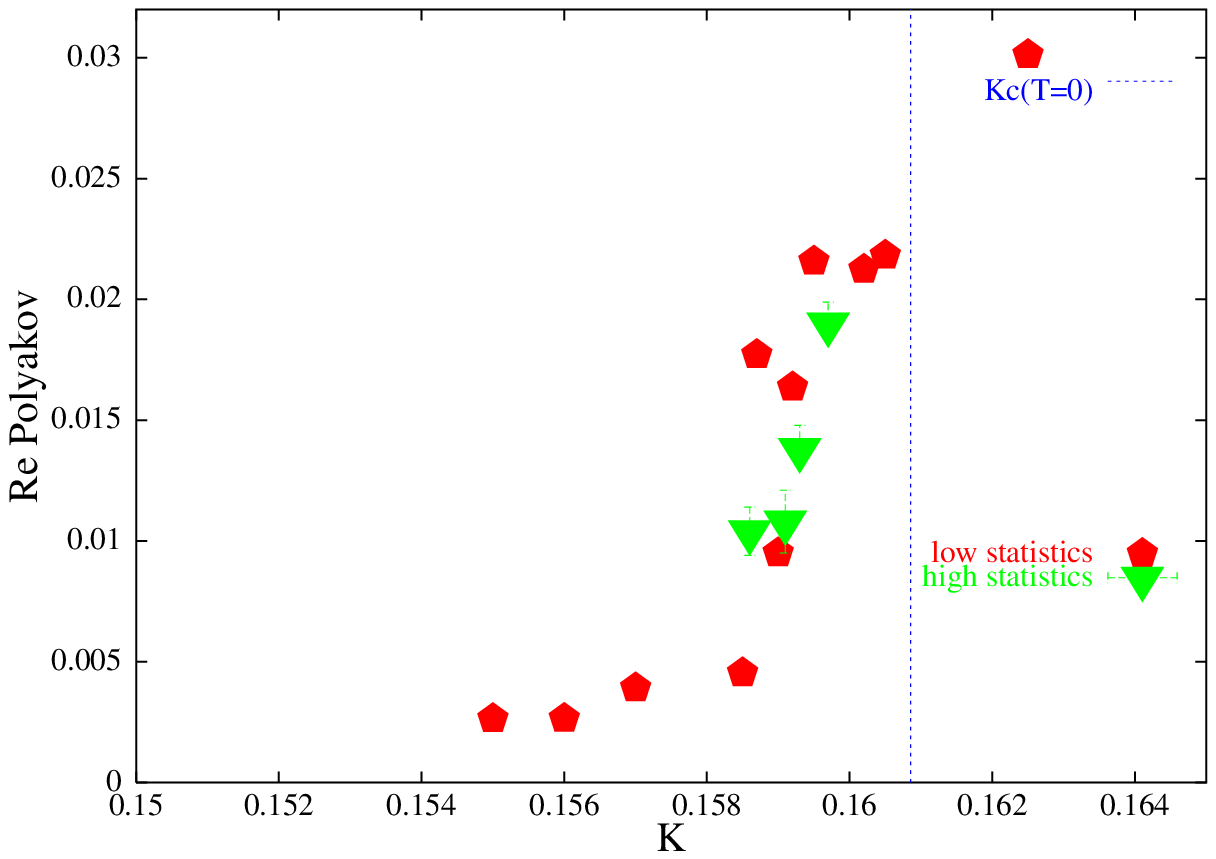}
\hspace*{0.5cm}
\includegraphics[width=7.0 truecm,height=5.5 truecm]{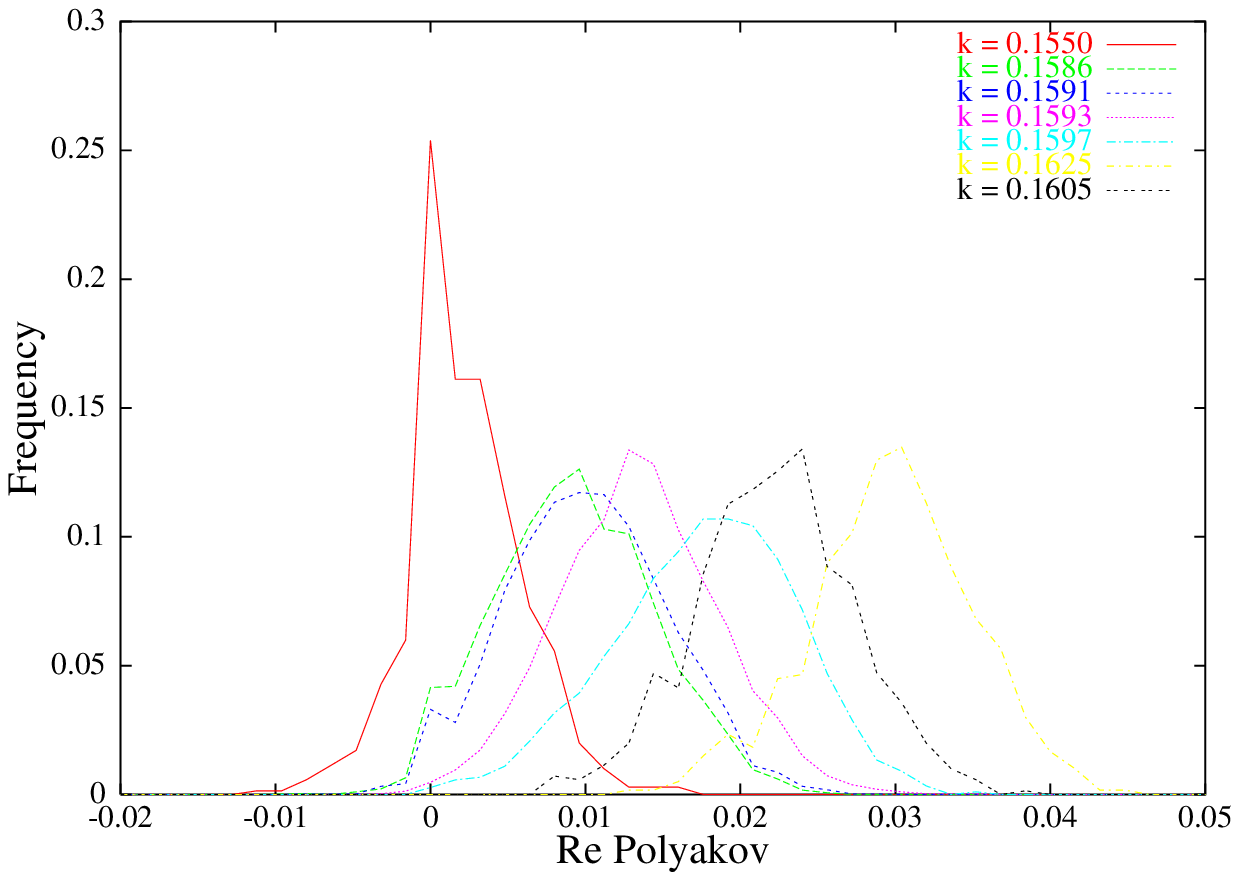}
\caption{The real part of the Polyakov loop as a function of $\kappa$
(left diagram), and the Polyakov loop histograms (right diagram)
at $\beta = 3.9$ and $\mu = 0.005$.
The data set is the same as in Figure 2, with the inclusion of
some high statistics results.
Both diagrams are consistent with a critical 
point or crossover at $\kappa_t = 0.1597(5)$. }
\end{figure}
The steepest slope of the plaquette and of the Polyakov loop, as well 
as the broadening of the probability distributions suggest a crossover 
or phase transition at $\kappa_t$:
\begin{equation}
\kappa_t(\beta = 3.9, \mu = 0.005) = 0.1597(5)
\end{equation}
and, as it was also observed at $\beta = 3.75$,
\begin{equation}
\kappa_t <  \kappa_c(T = 0) = 0.16085 \; .
\end{equation}

Although we postpone a detailed analysis of the spectrum and related observables
to our ongoing high statistics study~\cite{Progress}, we have 
collected a few results for the pion propagator.

The (zero momentum) pion propagator $G(t)$ is measured at a 
selected sample of couplings, 
and is fitted to a standard hyperbolic cosine form
\begin{equation}
G(t) = A~\cosh\left\{ M \left(t - \frac{N_t}{2} \right) \right\}
\end{equation}
in the time interval [2:6].
\begin{figure}[htb]
\includegraphics[width=7.0 truecm,height=5.5 truecm]{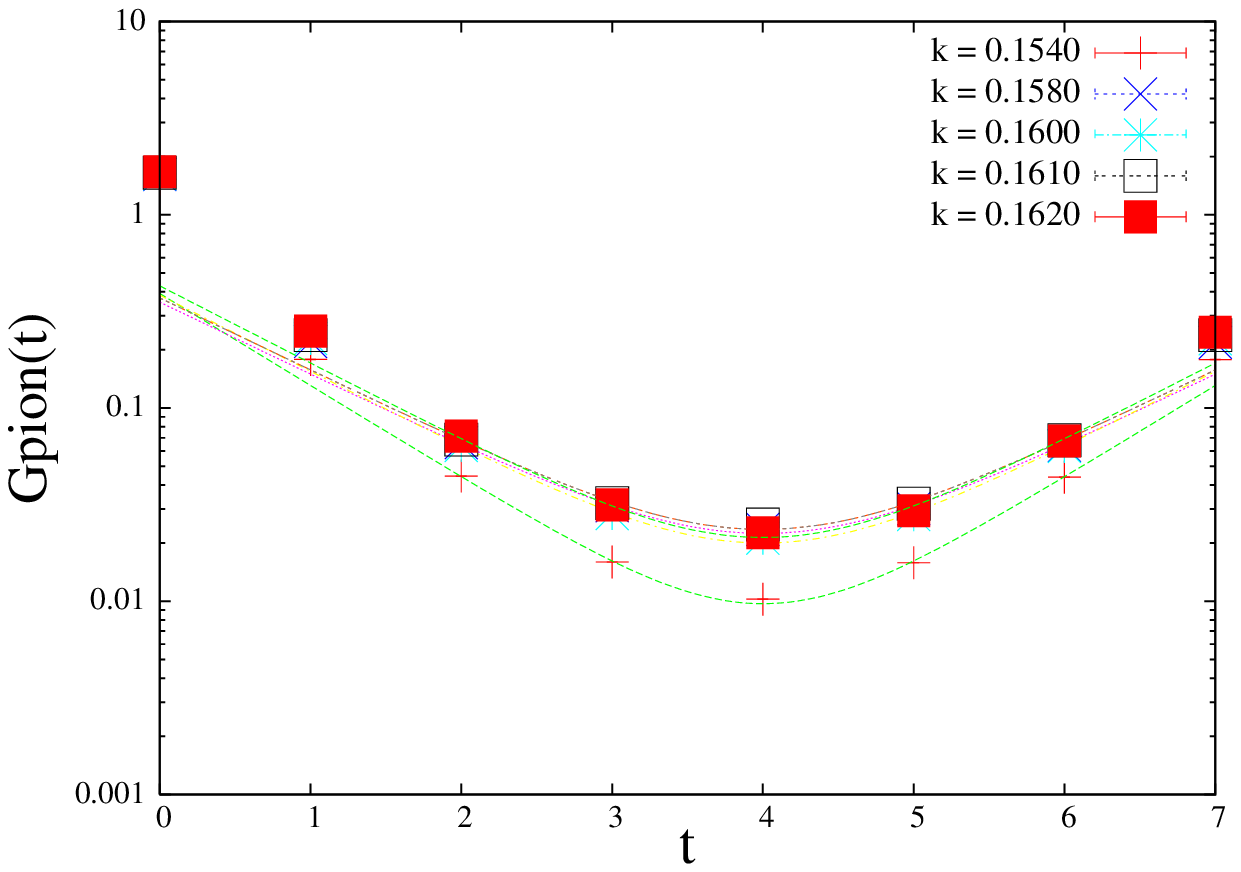}
\hspace*{0.5cm}
\includegraphics[width=7.0 truecm,height=5.5 truecm]{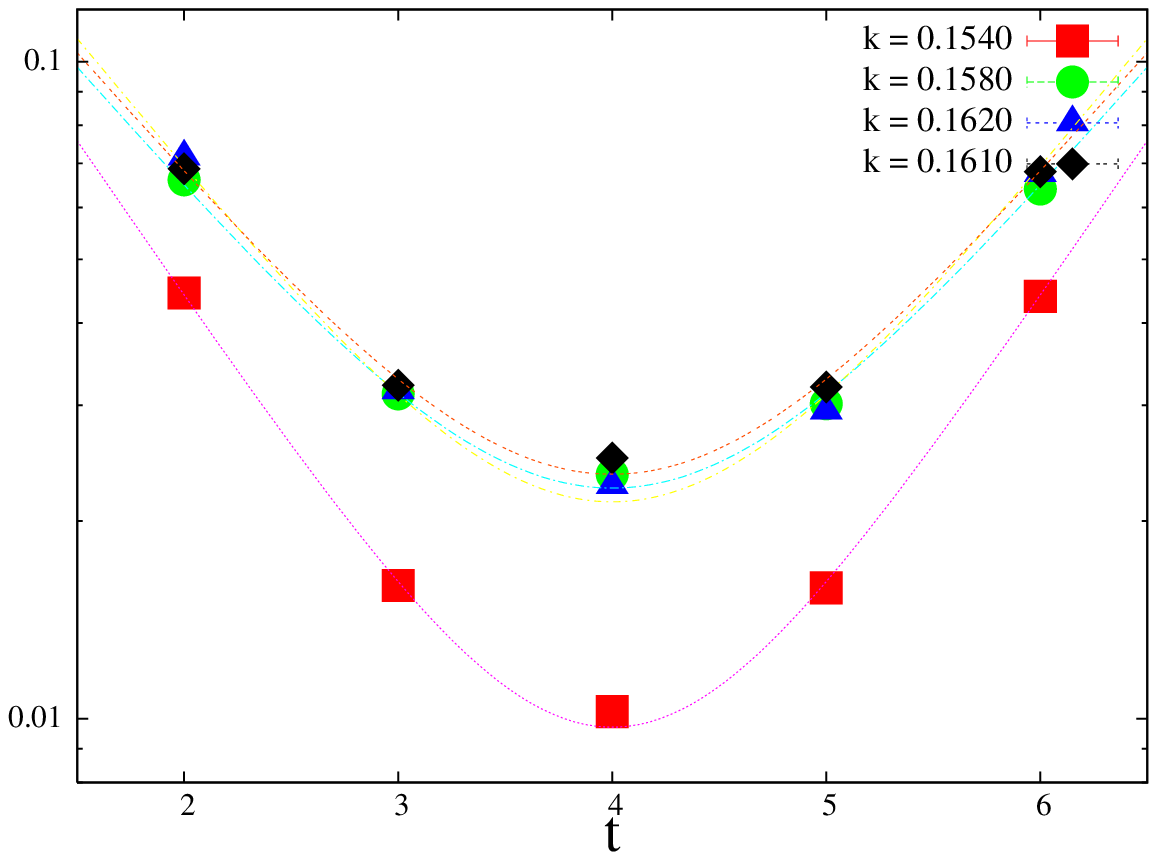}
\caption{The pion propagator $G(t)$ for selected $\kappa$ values. The 
solid lines are the simple fits described in the text. 
The right figure shows a subset of the same data points as in the left figure
on a different scale. }
\end{figure}
 
We show the quality of the results, with the fits themselves superimposed, 
in Fig. 6 (left and right, note a different scale between the two).
The resulting fit parameters, A(mplitude) and M(ass), 
are collected in Fig. 7 . 
They indicate that the effective pion mass decreases 
while approaching the thermal transition
from below, while the amplitude of the propagator increases, 
following the trend of the average plaquette.
\begin{figure}
\includegraphics[width=10.1 truecm,height=4.5 truecm]{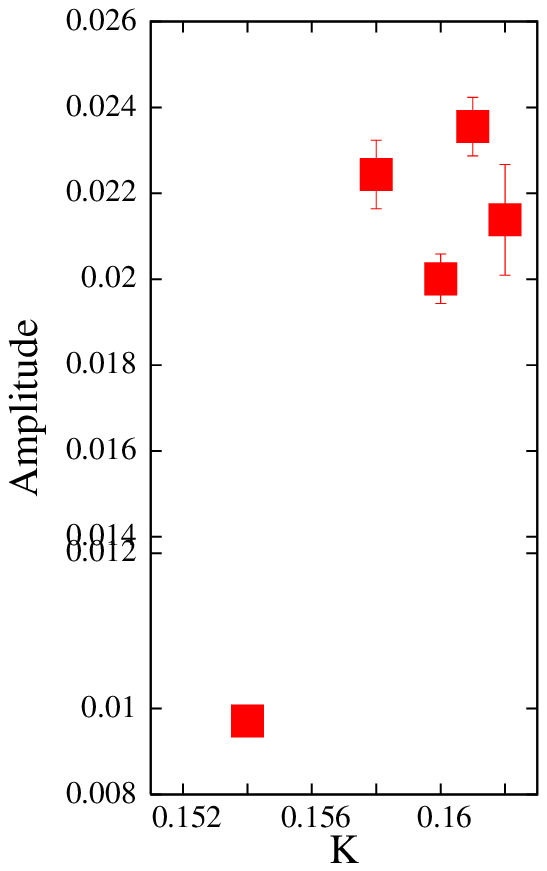}    
\hskip -1.0 truecm
\includegraphics[width=10.1 truecm,height=4.5 truecm]{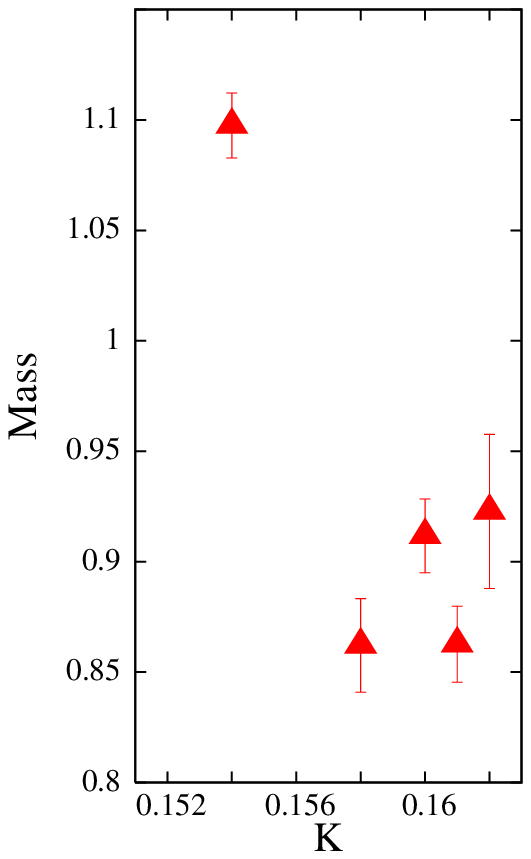}    
\caption{Amplitude of the propagator and effective pion mass  
from the hyperbolic cosine fits as a function of $\kappa$.} 
\end{figure}

\section{Summary and Outlook}

We have studied QCD with two flavors of dynamical
Wilson fermions including a twisted mass term on a $16^3\times 8$ lattice 
at two values of the temperature: $\beta = 3.75$ corresponding 
to $T \simeq 205$ MeV and $\beta = 3.9$ corresponding to 
$T \simeq  259$ MeV.

In either cases we have simulated $O(10)$ values of bare quark masses,
by varying the hopping parameter $\kappa$ at constant $\mu = 0.005$.
We have  observed a behavior consistent with a crossover 
(and not excluding a real transition) at a critical value of $\kappa$,
which we denoted as $\kappa_t$, which is less than the 
critical $\kappa$ at $T = 0$. This behavior is similar to that observed with
ordinary Wilson fermions~\cite{AliKhan:2000iz,AliKhan:2001ek}.

In our current study~\cite{Progress} on the apeNEXT machines at DESY and INFN   
we have to study next the $\mu$ dependence of our results. 
To this end, we are repeating a $\kappa$ scan at a larger $\mu$ value. 
At the same time we want to take full advantage of the 
twisted mass approach by working at full twist
with $\kappa = \kappa_c( \beta, T=0 )$. 
In this case the phase transition will be 
crossed by tuning the twisted mass $\mu$. Model studies along the lines
of Ref.~\cite{Creutz:1996bg}
will be most useful to guide our simulations.

\vskip .2 truecm
\noindent
{\bf Acknowledgments}:
It is a pleasure to thank Mike Creutz, Roberto Frezzotti, Agnes Mocsy,
GianCarlo Rossi and 
Carsten Urbach for interesting and helpful discussions. 
We wish also to thank the apeNEXT Collaboration, and in particular 
Alessandro Lonardo, Davide Rossetti, Hubert Simma, Raffaele Tripiccione and 
Piero Vicini, for their crucial help and support, as well as
Giampietro Tecchiolli for granting access to the Amaro apeNEXT prototype.

\end{document}